\documentclass[conference]{IEEEtran}
\usepackage[latin9]{inputenc}
\usepackage{amsmath}
\usepackage{amssymb}
\usepackage{graphicx}
\usepackage{fancyhdr}
\makeatletter
\IEEEoverridecommandlockouts
\usepackage{cite}
\usepackage{amsfonts}\usepackage{algorithmic}
\usepackage{subfigure}
\usepackage{textcomp}
\usepackage{xcolor}
\def\BibTeX{{\rm B\kern-.05em{\sc i\kern-.025em b}\kern-.08em
    T\kern-.1667em\lower.7ex\hbox{E}\kern-.125emX}}

\makeatother

\begin{document}

\title{Exceptional Point of Degeneracy in Linear-Beam Tubes for High Power
Backward-Wave Oscillators}

\author{%\thanks{Identify applicable funding agency here. If none, delete this.}
}

\author{\IEEEauthorblockN{Tarek Mealy, Ahmed F. Abdelshafy and Filippo Capolino}
\IEEEauthorblockA{\textit{Department of Electrical Engineering and Computer Science,
University of California, Irvine, CA 92697 USA} \\
 %\textit{University of California, Irvine, CA 92697 USA}\\
tmealy@uci.edu, abdelsha@uci.edu and f.capolino@uci.edu}}
\maketitle
\thispagestyle{fancy}
\begin{abstract}
An exceptional point of degeneracy (EPD) is induced in a system made
of an electron beam interacting with an electromagnetic (EM) guided
mode. This enables a degenerate synchronous regime in backward wave
oscillators (BWOs) where the electron beams provides distributed gain
to the EM mode with distributed power extraction. Current particle-in-cell
simulation results demonstrate that BWOs operating at an EPD have
a starting-oscillation current that scales quadratically to a non-vanishing
value for long interaction lengths and therefore have higher power
conversion efficiency at arbitrarily higher level of power generation
compared to standard BWOs.
\end{abstract}


\section{Introduction }

The characterizing feature of an exceptional point is the singularity
resulting from the degenracy of at least two eigenstates. We stress
the importance to refer to it as \textquotedblleft degeneracy\textquotedblright{}
as implied in \cite{berry2004physics}. Here an exceptional point
of degeneracy (EPD) is demonstrated in a system made of an electron
beam interacting with an electromagnetic (EM) guided mode. Despite
most of the published work on EPDs are related to parity time (PT)
symmetry \cite{bender1998real,klaiman2008visualization}, the occurrence
of EPDs does not necessarily require a system to exactly satisfy the
PT symmetry condition, however, it generally requires a system to
simultaneously have gain and loss \cite{liertzer2012pump}. The system
we consider in this paper involves two complete different media that
support waves: an electron beam (e-beam) for charge waves and a waveguide
for EM waves. Exchange of energy occurs when an EM waves in a slow
wave structure (SWS) interacts with the e-beam. In this paper the
degeneracy condition is enabled by the distributed power extraction
(DPE) from the SWS waveguide as shown in Fig. \ref{Fig:BWOs_All}.
The energy that is extracted from the e-beam and delivered to the
guided EM mode is considered as a distributed gain from the SWS prescriptive,
whereas the DPE represents extraction ``losses'' and not mere dissipation
\cite{mealy2019exceptional,mealy2019backward}. 

Backward-wave oscillators (BWOs) are high power sources where the
power is transferred from a very energetic e-beam to a synchronized
EM mode \cite{gilmour1994principles}. The extracted power in a conventional
BWO is usually taken at one end of the SWS \cite{levush1992theory,johnson1955backward}
as shown in Fig. \ref{Fig:BWOs_All}(a). One challenging issue in
BWOs is the limitation in power generation level. Indeed conventional
BWOs exhibit small starting beam current (to induce sustained oscillations)
and limited power efficiency without reaching very high output power
levels \cite{chen2000saturated}. Several techniques were proposed
in literature to enhance the power conversion efficiency of BWOs by
optimizing the SWS and its termination. For example, non-uniform SWSs
were proposed to enhance efficiency of BWOs in \cite{moreland1994efficiency},
in \cite{li2008investigation} a resonant reflector was used to enhance
efficiency to about 30\%, and a two-sectional SWS was also proposed
to enhance the power efficiency in \cite{zhang2009studies}. Here
we propose a regime of operation of BWO based on an EPD that to occurs
need a DPE as in Fig. \ref{Fig:BWOs_All}(b). In this paper we show
the physical mechanism of an EPD arising from the interaction of an
e-beam and and EM wave in a SWS and we show how this finding can be
used as a regime of operation in what we call an EPD-BWO to produce
very high power with high efficiency.

\section{Theoretical model based on an extension of the Pierce model}

\begin{figure}
\begin{centering}
\centering\includegraphics[width=0.8\columnwidth]{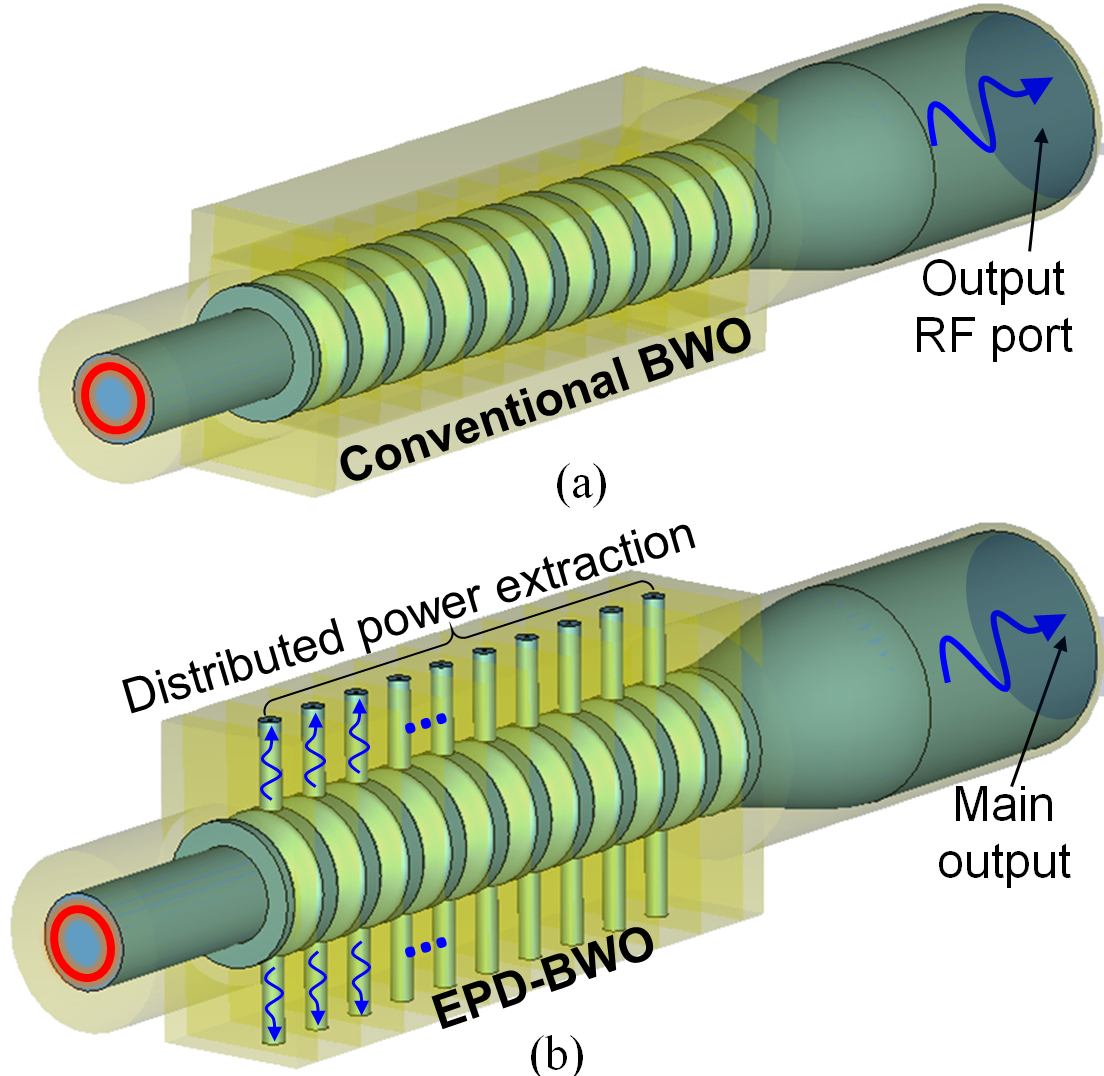} 
\par\end{centering}
\centering{}\caption{(a) Conventional BWO where the power is extracted from the waveguide
end; (b) EPD-BWO where the power is extracted in a distributed fashion
to satisfy the EPD condition. The power is extracted using distributed
wire loops (as an example) that are connected to coaxial waveguides.}
\label{Fig:BWOs_All}
\end{figure}

The interaction between the e-beam charge wave and the EM wave in
the SWS occurs when they are synchronized, i.e., when the EM wave
phase velocity $v_{ph}=\omega/\beta_{p}$ is matched to the average
velocity of the electrons $u_{0}$, where $\beta_{p}$ is the phase
propagation constants of the ``cold'' EM wave, i.e., when it is
not interacting with the e-beam. The synchronization condition provides
an estimate of the oscillation frequency of BWO ( $\omega\approx u_{0}\beta_{p}$
) and is considered as an initial criterion, because the phase velocity
of the ``hot'' modes, i.e., in the \textit{interactive} system,
are different from $v_{ph}$ and $u_{0}$ due to the interaction \cite{pierce1951waves,sturrock1958kinematics}.

The interaction between the e-beam and the EM wave in vacuum tube
devices was theoretically studied by Pierce in \cite{pierce1951waves}.
Assuming a wave eigenfunctions of the interactive system of infinite
length in the form of $\phi(z,t)\propto e^{i\omega t-ikz}$, Pierce
showed that the solutions of the linearized differential equations
that govern the electron beam charges' motion and continuity in presence
of the SWS EM field yield four eigenmodes whose dispersion relation
is given by the following characteristic equation \cite{pierce1951waves,sturrock1958kinematics}

\begin{equation}
\begin{array}{c}
D(\omega,k)=k^{4}-2\beta_{0}k^{3}+\left(\beta_{0}^{2}-\beta_{p}^{2}+\dfrac{I_{0}Z_{c}\beta_{p}\beta_{0}}{2V_{0}}\right)k^{2}\ \ \ \ \ \ \ \ \\
\ \ \ \ \ \ \ \ \ \ \ \ \ \ \ \ \ \ \ \ \ \ \ \ \ \ \ \ \ \ \ \ +2\beta_{0}\beta_{p}^{2}k-\beta_{0}^{2}\beta_{p}^{2}=0,
\end{array}\label{eq:Disp_1}
\end{equation}
where $\beta_{0}=\omega/u_{0}$ is the unmodulated beam wavenumber,
$V_{0}$ and $I_{0}$ are the e-beam equivalent dc voltage and dc
current, respectively, and $Z_{c}$ is characteristic impedance of
the cold EM mode. The Pierce model has been extended in Ref.\cite{mealy2019exceptional,mealy2019backward}
to the case of a SWS with DPE, where the propagation constant and
characteristic impedance of the cold EM mode are complex: $\beta_{p}=\beta_{pr}+i\beta_{pi}$
and $Z_{c}=Z_{cr}+iZ_{ci}$.

A second order EPD occurs in the interctive system when two solutions
of (\ref{eq:Disp_1}) are identical, $k_{1}=k_{2}=k_{e}$, where $k_{e}$
is the degenerate wavenumber, at a given angular frequency $\omega_{e}$.
This yields that \textit{two} hot modes have exactly the same phase
velocity $\omega/\mathrm{Re}(k_{e})$ which means that synchronization
is achieved in the interactive system and not in the cold system.
The conditions that lead to having two degenerate wavenumbers of hot
modes are $D(\omega_{e},k_{e})=0$ and $\partial_{k}D(\omega_{e},k)\big|_{k_{e}}=0$
\cite{hanson2018exceptional}, which yet is simplified by getting
rid of $k_{e}$ to (a detailed formulation of the derivation of the
following EPD condition is presented in Appendix \ref{sec:APP_A})

\begin{equation}
\left(\dfrac{\beta_{p}}{\beta_{0}}\right)^{2}=\left(\sqrt[3]{\dfrac{I_{0}Z_{c}\beta_{p}}{2V_{0}\beta_{0}}}+1\right)^{3}.\label{eq:EPD_Cond-1}
\end{equation}

Note that an EPD requires the coealscence of the two eigenvectors
associated to the two degenerate eigenvelaues as well. This has been
proven in Ref. \cite{mealy2019exceptional} by analytically determining
the two eigenvectors and by showing their analytical convergence.
Here we want to add another perspective to ensure the system has an
EPD, by showing that this strong degenerate condition is related to
the description of the two degenerate eigenvelues' perturbation in
terms of the Puiseux fractional power expansion \cite{welters2011explicit}
that, truncated to its first term, implies $(k_{n}-k_{e})\approx\left(-1\right)^{n}\alpha_{1}\sqrt{\omega-\omega_{e}}$
where $k_{n}$, with $n=1,2$, are the two perturbed wavenumbers in
the neighborhoos of $(\omega_{e},k_{e})$. The enabling factor for
this characterizing fractional power expansion is the fact that at
the point $(\omega_{e},k_{e})$ we have $\partial_{\omega}D(\omega,k_{e})\big|_{\omega_{e}}\neq0$
and therefore (\ref{eq:EPD_Cond-1}) will yield a branch point $(k-k_{e})\approx\alpha_{1}\sqrt{\omega-\omega_{e}}$
in the dispersion diagram, where $\alpha_{1}=\sqrt{-2\partial_{\omega}D/\partial_{k}^{2}D}\big|_{(\omega_{e},k_{e})}$
as shown in Ref. \cite{welters2011explicit}. We have verified that
this derivatives is indeed non vanishing in the presented cases. The
existance of the Puiseux series results in having a Jordan block in
the system matrix which is one of the characterizing features of EPDs,
as it was shown in \cite{mealy2019exceptional} in details, in terms
of the two coalescing eigenvectors.

The cold propagation constant imaginary part $\beta_{pi}$ accounts
for power attenuation along the SWS due to the leakage of power out
of the SWS as shown in Fig. \ref{Fig:BWOs_All}(b). Under the assumption
that $|\beta_{pi}|\ll|\beta_{pr}|$ the complex EPD condition in (\ref{eq:EPD_Cond-1})
is simplified to

\begin{equation}
\begin{array}{c}
I_{0}=I_{0e}\approx\dfrac{128}{81\sqrt{3}}\dfrac{V_{0}}{-Z_{cr}}\dfrac{\beta_{pi}^{3}}{\beta_{0}^{3}}\bigg|_{\beta_{pr}=\beta_{0}}.\end{array}\label{eq:EPD_Curr_Cond-1}
\end{equation}

A detailed formulation of the derivation and the assumptions used
to derive (\ref{eq:EPD_Curr_Cond-1}) is presented in Appendix \ref{sec:APP_B}.
From a theoretical perspective, the EPD condition is satisfied just
by tuning the e-beam dc current $I_{0}$ to a specific value which
we call EPD current $I_{0e}$ \cite{seyranian2005coupling}. The EPD
condition in (\ref{eq:EPD_Curr_Cond-1}) shows that the required e-beam
dc current $I_{0e}$ increases cubically when increasing the amount
of distributed extracted power, which is represented in terms of the
imaginary part $\beta_{pi}$ of the cold SWS's EM mode. The fact that
an EPD e-beam current $I_{0e}$ is found for any amount of distributed
power extraction, implies a tight (degenerate) synchronization regime
is guaranteed for any high power generation. Therefore, in principle
the synchronism is maintained for any desired distributed power output,
according to the Pierce-based model. Note that this trend is definitely
not observed in standard BWOs where interactive modes are non-degenerate
and the load is at one end of the SWS (i.e., $\beta_{pi}\approx0$
in SWSs made of copper without DPE).

The starting current for oscillation in a conventional BWO, where
the supported modes are non-degenerate, was theoretically studied
in \cite{johnson1955backward}. The \textit{starting oscillation condition}
is determined by imposing infinite gain $A_{v}\to\infty$ , where
the gain $A_{v}$ is defined as the field amplitude ratio at the begin
and end of the SWS \cite{johnson1955backward}. Accordingly , the
starting current of oscillation in a conventional BWO scales with
the SWS length $\ell$ as $I_{st}=\zeta/\ell^{3}$ \cite{walker1953starting,johnson1955backward}
, where $\zeta$ is a constant.

When a BWO with DPE operates in close proximity of the EPD, i.e.,
when the beam dc current $I_{0}$ is close to the EPD current $I_{0e}$,
there are two coalescing modes out of the three interacting modes
with positive $\mathrm{Re}(k)$ and they are denoted by $k_{1}=k_{e}+\alpha\sqrt{I_{0}-I_{0e}}$
and $k_{2}=k_{e}-\alpha\sqrt{I_{0}-I_{0e}}$ \cite{welters2011explicit},
where $\alpha=\sqrt{-2\partial_{I}D/\partial_{k}^{2}D}\big|_{(\omega_{e},k_{e})}$
is constant, whereas, the third mode is has $k_{3}=\left(\beta_{p}\beta_{0}^{2}\right)/\left(k_{1}k_{2}\right)$.
The gain expression for this case becomes \cite{mealy2019exceptional}

\begin{equation}
\begin{array}{c}
A_{ve}^{-1}e^{i\beta_{0}\ell}\approx\dfrac{e^{-i\left(k_{e}-\beta_{0}\right)\ell}\left(k_{e}-\beta_{0}\right)^{2}}{\left(k_{e}-k_{3}\right)}\dfrac{\sin\left(\alpha\sqrt{I_{0}-I_{0e}}\ell\right)}{\alpha\sqrt{I_{0}-I_{0e}}}\end{array}.\label{eq:Gain_Expr_Final}
\end{equation}
From (\ref{eq:Gain_Expr_Final}) we found that the oscillation condition
$A_{v}\to\infty$ , is verified when the beam dc current satisfies
$\alpha\sqrt{I_{0}-I_{0e}}=\pi/\ell$. Therfore the startig current
of oscillation is determined in term of the EPD current and the SWS
length as \cite{mealy2019exceptional}

\begin{equation}
\begin{array}{c}
I_{st}|_{EDP-BWO}=I_{0e}+\left(\dfrac{\pi}{\alpha\ell}\right)^{2}.\end{array}\label{eq:Ass_Scaling}
\end{equation}

\section{Application to overmoded SWS}

\begin{figure}
\begin{centering}
\centering \subfigure[]{\label{Fig:SWS_Conv}\includegraphics[width=0.48\columnwidth]{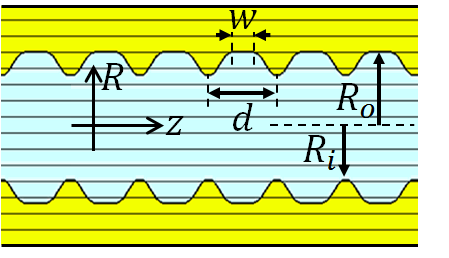}}\subfigure[]{\label{Fig:SWS_EPD_BWO}\includegraphics[width=0.48\columnwidth]{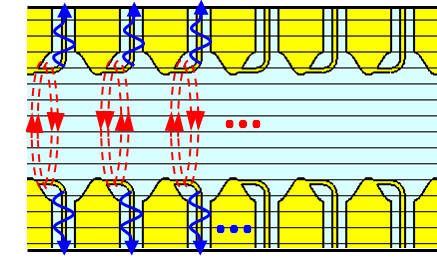}}
\par\end{centering}
\centering{}\caption{Longitudinal cross-sections of a SWS without (a) and with DPE (b).}
\end{figure}

\begin{figure}
\begin{centering}
\centering \subfigure[]{\label{Fig:Output_Conv}\includegraphics[width=0.98\columnwidth]{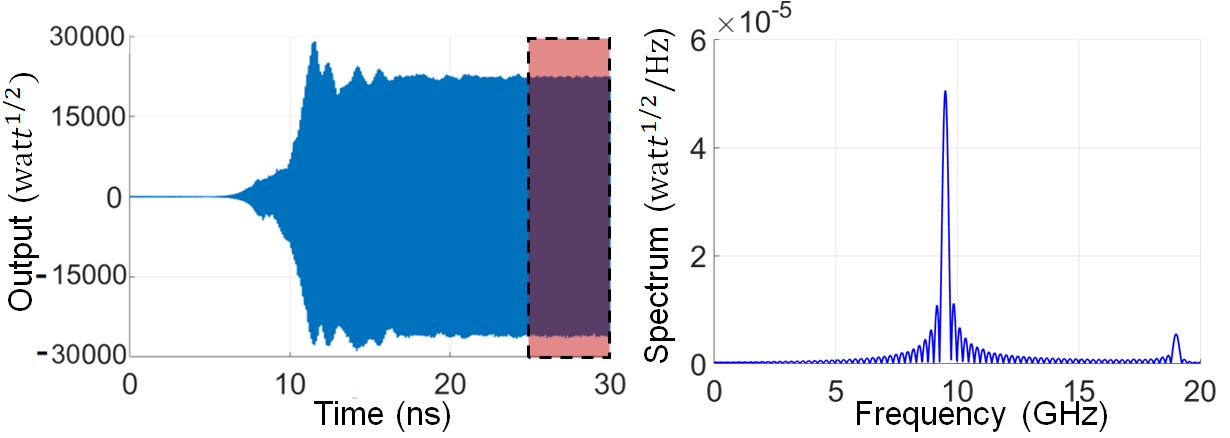}}
\par\end{centering}
\begin{centering}
\centering \subfigure[]{\label{Fig:Output_EPD}\includegraphics[width=0.98\columnwidth]{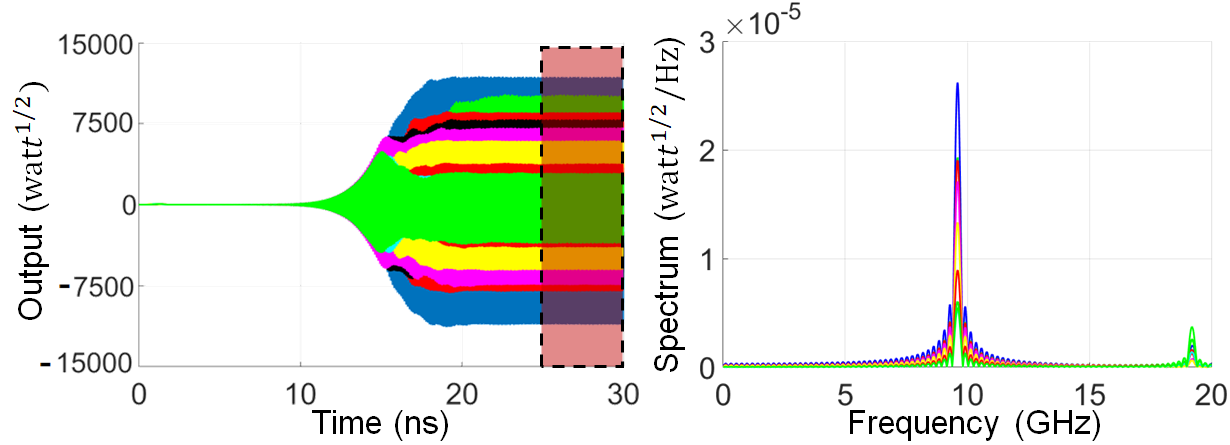}}
\par\end{centering}
\centering{}\caption{Output signals and their corresponding spectra for: (a) Conventional
BWO where the output power is only extracted from one port as shown
in Fig. \ref{Fig:BWOs_All}(a). (b) EPD-BWO where power is extracted
from multiple ports as shown in Fig. \ref{Fig:BWOs_All}(b).}
\label{Fig:Output_Vs_Time}
\end{figure}

As a proof of concept, we demonstrate the EPD-BWO regime by considering
a conventional BWO operating at X-band whose SWS is shown in Fig.
2(a). The DPE is introduced by adding two wire loops in each unit
cell, above and below as shown in Fig. \ref{Fig:SWS_EPD_BWO}, that
couple to the azimuthal magnetic field and by Farady's Law an electromotive
force is generated that excites each coaxial waveguide, similarly
to the way power is extracted from magnetrons (Ch. 10 in Ref. \cite{gilmour1994principles}).
Simulations based on the particle in cell solver (PIC), implemented
in CST Studio Suite, use a relativistic annular e-beam with dc voltage
$V_{0}=600$ kV. The output signals and their corresponding spectra
for both BWOs, with and without DPE, are shown in Fig. \ref{Fig:Output_Vs_Time}
where a self-standing oscillation frequency of 9.7 GHz is observed
when the used beam dc current is $I_{0}=1740$ A for both cases. Details
about the structure geometry and simulation settings are presented
in Appendix \ref{sec:APP_C}.

\begin{figure}
\begin{centering}
\label{Fig:Starting_Current}\includegraphics[width=0.99\columnwidth]{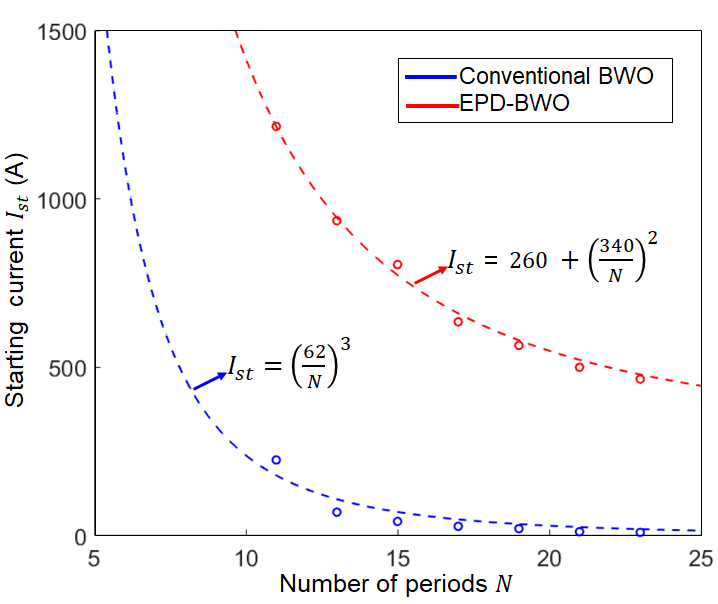}
\par\end{centering}
\centering{}\caption{Scaling of starting e-beam current for oscillation in conventional
BWO and EPD-BWO. Dashed lines represent fitting curves. The EPD-BWO
shows a starting current trend that does not vanish for long SWS.}
\end{figure}

To assess the occurrence of an EPD we verify the unique scaling trend
of the starting current in (\ref{eq:Ass_Scaling}). Fig. \ref{Fig:Starting_Current}
shows the starting current scaling trends for both conventional BWO
and EPD-BWO based on PIC simulation results. The dashed lines represent
fitting curves and the case of EPD-BWO shows very good fitting with
99\% R-square. In comparison to a conventional BWO, the EPD-BWO is
characterized by a starting current (threshold) that does not tend
to zero as the SWS length increases, and a scaling that is a quadratic
function of the inverse of the SWS length. The procedure to determine
the starting current of oscillation using PIC simulations is presented
in Appendix \ref{sec:APP_C}.

\begin{figure}
\begin{centering}
\label{Fig:Eff_Comparison}\includegraphics[width=0.99\columnwidth]{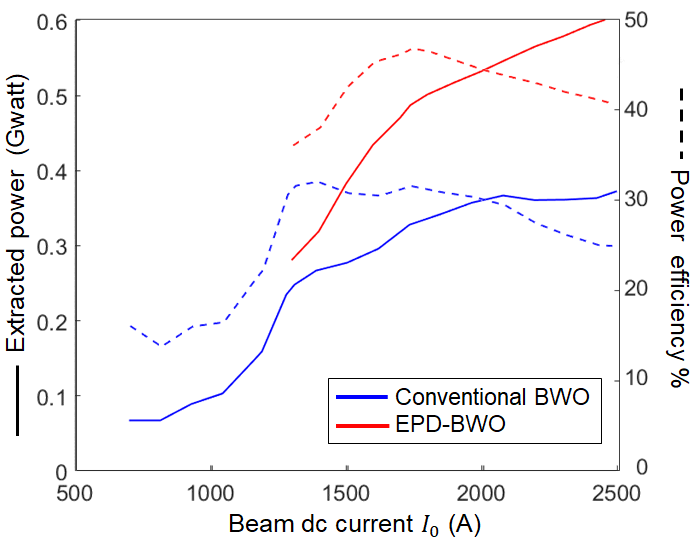}
\par\end{centering}
\centering{}\caption{Comparison between the efficiency of a conventional BWO and an EPD-BWO.
The EPD-BWO shows improved efficiency at higher level of power generation
compared to the conventional BWO.}
\end{figure}

We compare the RF conversion power efficiency (RF output power over
dc e-beam power) of the conventional BWO with that of the EPD-BWO
in Fig. \ref{Fig:Eff_Comparison} for e-beam dc currents that exceed
the starting current, assuming the SWS has 11 unit-cells. The figure
shows that the EPD-BWO has higher efficiency at higher level of output
power compared to a conventional BWO with same dimensions. The results
show that the EPD-BWO has a maximum efficiency of about 47\% at about
0.5 GW output power (the sum of the power from each output in Fig.
\ref{Fig:BWOs_All}(b)). Instead, the conventional BWO has a maximum
efficiency of about 33\% at an output power level of about 0.27 GW.
It is important to point out that the EPD-BWO has a higher threshold
beam current to start oscillations compared to the conventional one
which is in consistent with the theoretical results in \cite{mealy2019exceptional}
and with the requirement of generating higher power levels.

\section{Conclusion}

In summary, the physical mechanism of an EPD in a hybrid system where
a linear electron beam interacts with an electromagnetic mode has
been demonstrated. The manifestation of such EPD is useful to conceive
a new degenerate synchronous regime for BWOs that have a starting-oscillation
current law that decreases quadratically to a given fixed value for
long waveguide interaction lengths; as a consequence PIC simulations
show higher efficiency and much higher output power than a standard
BWO. The unique quadratic threshold scaling law demonstrates the EPD-based
synchronization phenomenon compared to that in a standard BWO that
has a starting-oscillation current law that vanishes cubically for
long waveguide interaction lengths.

\section{Acknowledgment}

This material is based upon work supported by the Air Force Office
of Scientific Research award number FA9550-18-1-0355. The authors
are thankful to DS SIMULIA for providing CST Studio Suite that was
instrumental in this study.

\appendices{}

\numberwithin{equation}{section}

\section{Second order EPD in a system made of an electromagnetic wave interacting
with an electron beam's charge wave\label{sec:APP_A}}

The interaction between an electron (e)-beam and an electromagnetic
(EM) wave in a linear vacuum tube was theoretically studied by Pierce
in \cite{pierce1951waves}. Assuming wave eigenfunctions of the interactive
system of infinite length in the form of $\phi(z,t)\propto e^{i\omega t-ikz}$,
Pierce showed that the solutions of the linearized differential equations
that govern the electron beam charges' motion and continuity in presence
of the slow wave structure (SWS) EM field yield four eigenmodes whose
wavenumber dispersion is given by the characteristic equation \cite{pierce1951waves,sturrock1958kinematics}

\begin{equation}
\begin{array}{c}
D(\omega,k)=k^{4}-2\beta_{0}k^{3}+\left(\beta_{0}^{2}-\beta_{p}^{2}+\dfrac{I_{0}Z_{c}\beta_{p}\beta_{0}}{2V_{0}}\right)k^{2}\ \ \ \ \ \ \ \ \\
\ \ \ \ \ \ \ \ \ \ \ \ \ \ \ \ \ \ \ \ \ \ \ \ \ \ \ \ \ \ \ \ +2\beta_{0}\beta_{p}^{2}k-\beta_{0}^{2}\beta_{p}^{2}=0,
\end{array}\label{eq:Disp_1-1}
\end{equation}
Here $\beta_{0}=\omega/u_{0}$ is the unmodulated e-beam wavenumber,
$u_{0}$ is the average velocity of the electrons, $V_{0}$ and $I_{0}$
are the e-beam equivalent dc voltage and dc current, respectively,
and $\beta_{p}$ and $Z_{c}$ are the wavenumber and characteristic
impedance, respectively, of the EM mode in the ``cold'' (i.e., without
interacting with the e-beam) SWS. The dispersion equation in (\ref{eq:Disp_1-1})
has four root solutions. The four complex wavenumbers are the eigenvalues
of the interactive (also called ``hot'') system. A necessary condition
to have a second order exceptional point of degeneracy (EPD) is to
have two repeated eigenvalues, which means that at the EPD frequency
$\omega=\omega_{e}$ the characteristic equation should have two repeated
roots as

\begin{equation}
\begin{array}{c}
D(\omega_{e},k)\propto(k-k_{e})^{2}\end{array}\label{eq:Disp_Req}
\end{equation}
where $k_{e}$ is the degenerate wavenumber. The relation in (\ref{eq:Disp_Req})
is satisfied when \cite{hanson2018exceptional}

\begin{equation}
\begin{array}{c}
D(\omega_{e},k_{e})=0,\\
\partial_{k}D(\omega_{e},k)\Big|_{k=k_{e}}=0.
\end{array}\label{eq:EPD_Con1}
\end{equation}

Using the expression for $D(\omega,k)$ in (\ref{eq:Disp_1-1}), the
two necessary EPD conditions in (\ref{eq:EPD_Con1}) are explicitly
written as

\begin{equation}
\begin{array}{c}
k_{e}^{4}-2\beta_{0}k_{e}^{3}+\left(\beta_{0}^{2}-\beta_{p}^{2}+\dfrac{I_{0}Z_{c}\beta_{p}\beta_{0}}{2V_{0}}\right)k_{e}^{2}\ \ \ \ \ \ \ \ \\
\ \ \ \ \ \ \ \ \ \ \ \ \ \ \ \ \ \ \ \ \ \ \ \ \ \ \ \ \ \ \ \ +2\beta_{0}\beta_{p}^{2}k_{e}-\beta_{0}^{2}\beta_{p}^{2}=0,
\end{array}\label{eq:Cond1}
\end{equation}

\begin{equation}
4k_{e}^{3}-6\beta_{0}k_{e}^{2}+2\left(\beta_{0}^{2}-\beta_{p}^{2}+\dfrac{I_{0}Z_{c}\beta_{p}\beta_{0}}{2V_{0}}\right)k_{e}+2\beta_{0}\beta_{p}^{2}=0.\label{eq:Cond2}
\end{equation}

Note that the frequency dependency is in the terms $\beta_{0}=\omega/u_{0}$,
$\beta_{p}$ and $Z_{c}$. The frequency dependency in the cold SWS
terms $\beta_{p}$ and $Z_{c}$ depends on the waveguide geometry
and they can be approximated using simple distributed circuit models
in the neighborhood of the operative frequency. Using a simple transmission
line circuit model that supports \textit{backward propagation}, the
distributed per-unit-length series impedance $Z$ and shunt admittance
$Y$ are $\beta_{p}=\sqrt{-ZY}$ and $Z_{c}=\sqrt{Z/Y}$, respectively,
as also discussed in Ref. \cite{mealy2019exceptional}. The two square
roots values of each physical parameter represent waves that propagate
in opposite directions in the cold SWS, i.e., both $\beta_{p}$ and
$-\beta_{p}$ are valid solutions because of reciprocity, where the
root $\beta_{p}=\sqrt{-ZY}$ is taken as the principle square root
(resulting in a positive real part) while the root of $Z_{c}=\sqrt{Z/Y}$
is taken as the secondary square root (resulting in a negative real
part) as discussed in \cite{ziolkowski2001wave,mealy2019exceptional}.

We simplify the above two equations by first getting $Z_{c}\beta_{p}$
from (\ref{eq:Cond2}) as

\begin{equation}
Z_{c}\beta_{p}=\dfrac{2V_{0}\left(k_{e}-\beta_{0}\right)\left(\beta_{p}^{2}+k_{e}\left(\beta_{0}-2k_{e}\right)\right)}{I_{0}\beta_{0}k_{e}},\label{eq:Cond3}
\end{equation}
which is then used in (\ref{eq:Cond1}) to get $\beta_{p}^{2}$ as

\begin{equation}
\beta_{p}^{2}=\dfrac{k_{e}^{3}}{\beta_{0}}.\label{eq:Cond4}
\end{equation}
We then simplify (\ref{eq:Cond3}) by inserting (\ref{eq:Cond4})
in its right hand side to get

\begin{equation}
Z_{c}\beta_{p}=\dfrac{2V_{0}\left(k_{e}-\beta_{0}\right)^{3}}{I_{0}\beta_{0}^{2}}.\label{eq:Cond5}
\end{equation}

The conditions in (\ref{eq:Cond4}) and (\ref{eq:Cond5}) are constraints
on the cold SWS's circuit parameters $\beta_{p}$ and $Z_{c}$ to
enable the an EPD, and are important to select what kind of SWS shall
be chosen to ensure an EPD occurs at a given $(\omega_{e},k_{e})$.

The above equations can also be used to determine the EPD wavenumber
of the hot SWS, i.e., the wavenumber of the degenerate mode in the
interactive e-beam-EM system; from (\ref{eq:Cond4}) one obtains
\begin{equation}
k_{e}=\sqrt[3]{\beta_{p}^{2}\beta_{0}}.\label{eq:EPD_ke}
\end{equation}
 Then, using this $k_{e}$ expression in (\ref{eq:Cond5}), we obtain

\begin{equation}
\left(\dfrac{\beta_{p}}{\beta_{0}}\right)^{2}=\left(\sqrt[3]{\dfrac{I_{0}Z_{c}\beta_{p}}{2V_{0}\beta_{0}}}+1\right)^{3}.\label{eq:EPD_Cond-1-1}
\end{equation}

The above condition represents a constraint involving the operational
frequency $\omega$, electron beam dc voltage$V_{0}$ and current
$I_{0}$, and cold SWS circuit wavenumber $\beta_{p}$ and characteristic
impedance $Z_{c}$ to have an EPD. 

When distributed power extraction (DPE) occurs in the SWS, the propagation
constant and characteristic impedance of the ``cold'' EM mode (i.e.,
without coupling to the electron beam) are complex: $\beta_{p}=\beta_{pr}+i\beta_{pi}$
and $Z_{c}=Z_{cr}+iZ_{ci}$. The cold propagation constant imaginary
part $\beta_{pi}$ accounts for power attenuation along the SWS due
to the leakage of power out of the SWS. Note that $\beta_{pr}\beta_{pi}>0$
for a ``backward'' EM wave that is traveling in the cold SWS (we
are using the $\exp(i\omega t)$ time dependency which implies that
the EM modes propagates as $\exp(-i\beta_{p}z)$). Since the phase
propagation constant $\beta_{pr}$ is positive, because it has to
matche the electron beam effective wavenumber $\beta_{0}=\omega/u_{0}$,
one has $\beta_{pi}>0$. Furthermore, for a backward wave with $\beta_{pr}>0$,
one has $Z_{cr}<0$ since power travels along the $-z$ direction
in the cold SWS. Therefore in the above formulas we have that $Z_{c}\beta_{p}=\left(Z_{cr}\beta_{pr}-Z_{ci}\beta_{pi}\right)+i\left(Z_{ci}\beta_{pr}+Z_{cr}\beta_{pi}\right)$
is complex.

\begin{figure}
\begin{centering}
\centering \subfigure[]{\label{Fig:BWO_NPE}\includegraphics[width=1\columnwidth]{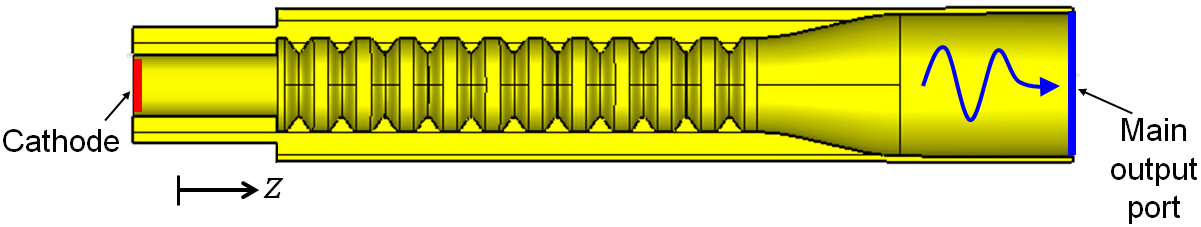}}
\par\end{centering}
\begin{centering}
\centering\subfigure[]{\label{Fig:BWO_DPE}\includegraphics[width=1\columnwidth]{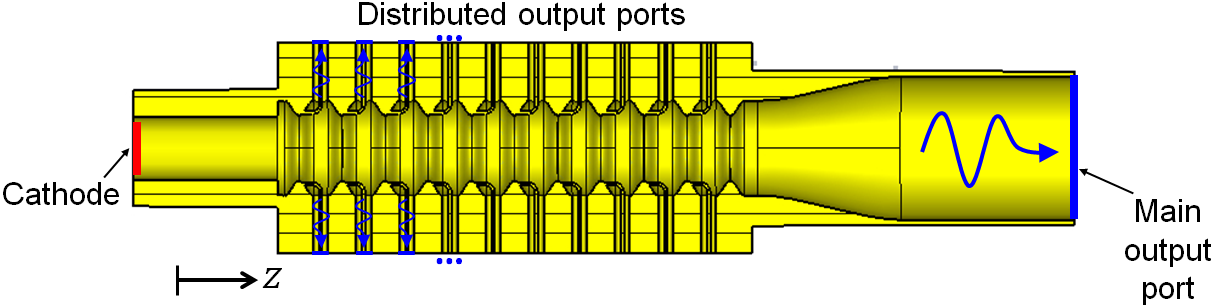}}
\par\end{centering}
\begin{centering}
\caption{Longitudinal cross-sections of (a) the conventional BWO considered
in this paper where the output power is extracted from the waveguide
right end; (b) the EPD-BWO where the power is extracted in a distributed
fashion to satisfy the EPD condition. The power is extracted using
distributed wire loops (as an example) that are connected to coaxial
waveguides on the top and bottom sides of the waveguide. This is just
an example of DPE to provide the proof of concept of an EPD in a BWO;
many other geometries are possible and the physical mechanism would
be analogous to the one demosntrated in this paper.}
\par\end{centering}
\centering{}\label{Fig:BWOs}
\end{figure}

\section{Electron beam dc current that satisfies the EPD condition\label{sec:APP_B}}

The electron beam current that satisfies the EPD condition is determined
by rearranging (\ref{eq:EPD_Cond-1-1}) as

\begin{equation}
I_{0}=I_{0e}\equiv\dfrac{2V_{0}\beta_{0}}{Z_{c}\beta_{p}}\left(\left(\dfrac{\beta_{p}}{\beta_{0}}\right)^{2/3}-1\right)^{3}.\label{eq:EPD_Cond-1-2}
\end{equation}

To satisfy the above condition, since the e-beam dc current $I_{0}$
is real valued, the imaginary part of the right hand side should vanish,
i.e.,

\begin{equation}
\begin{array}{c}
\mathrm{arg}\left(\dfrac{2V_{0}\beta_{0}}{Z_{c}\beta_{p}}\left(\left(\dfrac{\beta_{p}}{\beta_{0}}\right)^{2/3}-1\right)^{3}\right)=2n\pi,\end{array}n=\{0,\pm1,..\}.\label{eq:Arg_Cond}
\end{equation}

The propagation constant and characteristic impedance of the backward
EM mode are complex, and the imaginary part $\beta_{pi}>0$ of the
cold propagation constant accounts for distributed power extraction.
Under the assumption that $0<\beta_{pi}\ll\beta_{pr}$ and $|Z_{ci}|\ll|Z_{cr}|$
and considering a backward propagating mode so that $\mathrm{Re}\left(Z_{c}\beta_{p}\right)<0$,
it can be easily shown that $\left|\mathrm{Re}\left(Z_{c}\beta_{p}\right)\right|>\left|\mathrm{Im}\left(Z_{c}\beta_{p}\right)\right|$.

By assuming that the EPD point at $(\omega,k)=(\omega_{e},k_{e})$
is close to the synchronization point of the non interactive diagrams
(that is $(\omega,\beta_{p})\approx(\omega,\beta_{0})$) , i.e., we
impose that at $\omega=\omega_{e}$one has $\beta_{p}=\beta_{0}(1+\delta)$,
where $\delta=\delta_{r}+i\delta_{i}$, and $\delta_{i}>0$ (because
of losses and (DPE) in the cold SWS supporting the backward mode).
Because we assume that both $|\delta_{r}|\ll1$ and $\delta_{i}\ll1$,
the argument of the complex value in (\ref{eq:Arg_Cond}) is dominated
by the latter term, i.e.,

\begin{equation}
\begin{array}{c}
\mathrm{arg}\left(\dfrac{2V_{0}\beta_{0}}{Z_{c}\beta_{p}}\left(\left(\dfrac{\beta_{p}}{\beta_{0}}\right)^{2/3}-1\right)^{3}\right)\ \ \ \ \ \ \ \ \\
\ \ \ \ \ \ \ \ \ \ \ \ \ \ \ \ \ \ \ \ \ \ \ \ \ \ \ \ \ \ \ \ \approx\mathrm{\pi+3arg}\left(\left(\dfrac{\beta_{p}}{\beta_{0}}\right)^{2/3}-1\right).
\end{array}\label{eq:Arg_with_cubic_Root}
\end{equation}

The cubic root in (\ref{eq:Arg_with_cubic_Root}) has three solutions:

\begin{equation}
\begin{array}{cc}
\left(\dfrac{\beta_{p}}{\beta_{0}}\right)^{2/3}\approx\left(1+\dfrac{2}{3}\delta\right)e^{i2m\pi/3}, & m=\{0,1,2\}.\end{array}\label{eq:final_angle_cond-1}
\end{equation}

Considering the cubic root solution with $m=0$, the argument in (\ref{eq:Arg_with_cubic_Root})
is simplified to

\begin{equation}
\begin{array}{c}
\mathrm{arg}\left(\dfrac{2V_{0}\beta_{0}}{Z_{c}\beta_{p}}\left(\left(\dfrac{\beta_{p}}{\beta_{0}}\right)^{2/3}-1\right)^{3}\right)\ \ \ \ \ \ \ \ \\
\ \ \ \ \ \ \ \ \ \ \ \ \ \ \ \ \ \ \ \ \ \ \ \ \ \ \ \ \ \ \ \ \approx\mathrm{\pi+3arg}\left(\dfrac{2}{3}\delta\right)\\
\ \ \ \ \ \ \ \ \ \ \ \ \ \ \ \ \ \ \ \ \ \ \ \ \ \ \ \ \ \ \ \ =\pi+3\tan^{-1}\left(\dfrac{\delta_{i}}{\delta_{r}}\right).
\end{array}\label{eq:Arg_simplified}
\end{equation}

By enforcing angle condition in (\ref{eq:Arg_Cond}) to (\ref{eq:Arg_simplified})
we obtain

\begin{equation}
\begin{array}{cc}
\pi+3\tan^{-1}\left(\dfrac{\delta_{i}}{\delta_{r}}\right)=2n\pi, & n=0,\pm1,...\end{array}\label{eq:Final_Arg_eq}
\end{equation}

A relation between $\delta_{r}$ and $\delta_{i}$ is determined by
solving (\ref{eq:Final_Arg_eq}) which finally yields three possible
solutions

\begin{equation}
\delta_{i}=\begin{cases}
\begin{array}{c}
0\\
\sqrt{3}\delta_{r}\\
-\sqrt{3}\delta_{r}
\end{array}\end{cases}.\label{eq:final_angle_cond}
\end{equation}

We neglect the solution $\delta_{i}=0$ in (\ref{eq:final_angle_cond})
because the regime we are considering has DPE which implies that $\delta_{i}>0.$
Using the solution $\delta_{i}=\pm\sqrt{3}\delta_{r}$ in (\ref{eq:EPD_Cond-1-2})
will finally find the EPD current to be

\begin{equation}
I_{0e}\approx\dfrac{2V_{0}}{Z_{cr}}\left(\dfrac{2}{3}\left(\delta_{r}\pm i\sqrt{3}\delta_{r}\right)\right)^{3}=\dfrac{128}{81\sqrt{3}}\dfrac{V_{0}\delta_{i}^{3}}{(-Z_{cr})}
\end{equation}

Therefore, the EPD condition is met by just tuning the e-beam dc current
$I_{0}$ to a specific value which we call EPD e-beam current $I_{0e}$:

\begin{equation}
I_{0}=I_{0e}\approx\dfrac{128}{81\sqrt{3}}\dfrac{V_{0}}{\left(-Z_{cr}\right)}\dfrac{\beta_{pi}^{3}}{\beta_{0}^{3}}\bigg|_{\beta_{pr}=\beta_{0}}.\label{eq:EPD_ebeam_current}
\end{equation}

The other two solutions of the cubic root in (\ref{eq:final_angle_cond-1})
with $m=1$ and $m=2$ are discarded because they provide solutions
for a purely real right hand side of Eq. (\ref{eq:EPD_Cond-1-2})
for $|\delta_{i}|>1$ and $\left|\delta_{r}\right|>1$, that contradict
the initial assumption of $|\delta_{r}|\ll1$ and $\delta_{i}\ll1$.
In summary, the EPD occurs when the e-beam current $I_{0}$ takes
the value in (\ref{eq:EPD_ebeam_current}).

\section{Full-wave simulations' details\label{sec:APP_C}}

We demonstrate the EPD-BWO regime by taking a conventional BWO design
operating at X-band shown in Fig. \ref{Fig:BWOs}(a). The proposed
EPD-BWO is shown is Fig. \ref{Fig:BWOs}(b) where DPE is introduced
using distributed wire loops that are connected to coaxial waveguides.
The original SWS geometry shown in Fig. \ref{Fig:SWS_Conv-1}, is
a circular copper waveguide with azimuthal symmetry and with inner
and outer radii of $R_{i}$ = 11.5 mm and $R_{o}=16.5$ mm, respectively,
and period $d=15$ mm. The surface corrugation of SWS in one period
is described by a flat surface $R(z)=R_{o}$ for $0\leq z<w$, where
$w=5$ mm, and a sinusoidal corrugated surfaces for the rest of the
period described as $R(z)=(R_{o}+R_{i})/2+((R_{o}-R_{i})/2)\cos(2\pi(z-w)/(d-w))$
for $w\leq z<d$. The whole body of the BWO is made of copper with
vacuum inside.

\subsection{Cold simulations - EM modes in the cold SWS}

\begin{figure}
\begin{centering}
\centering \subfigure[]{\label{Fig:SWS_Conv-1}\includegraphics[width=0.48\columnwidth]{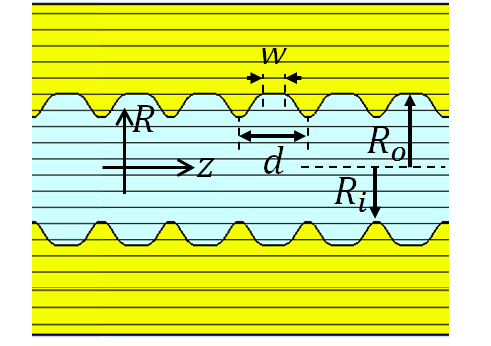}}\subfigure[]{\label{Fig:SWS_EPD_BWO-1}\includegraphics[width=0.48\columnwidth]{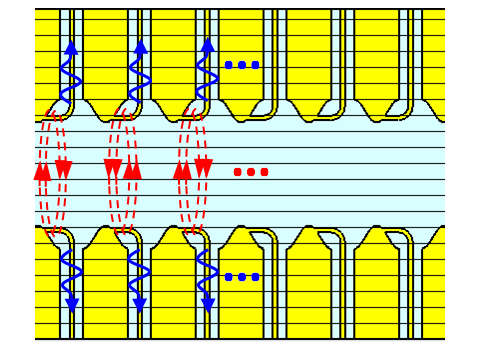}} 
\par\end{centering}
\begin{centering}
\centering \subfigure[]{\label{Fig:Cold_Disp}\includegraphics[width=0.95\columnwidth]{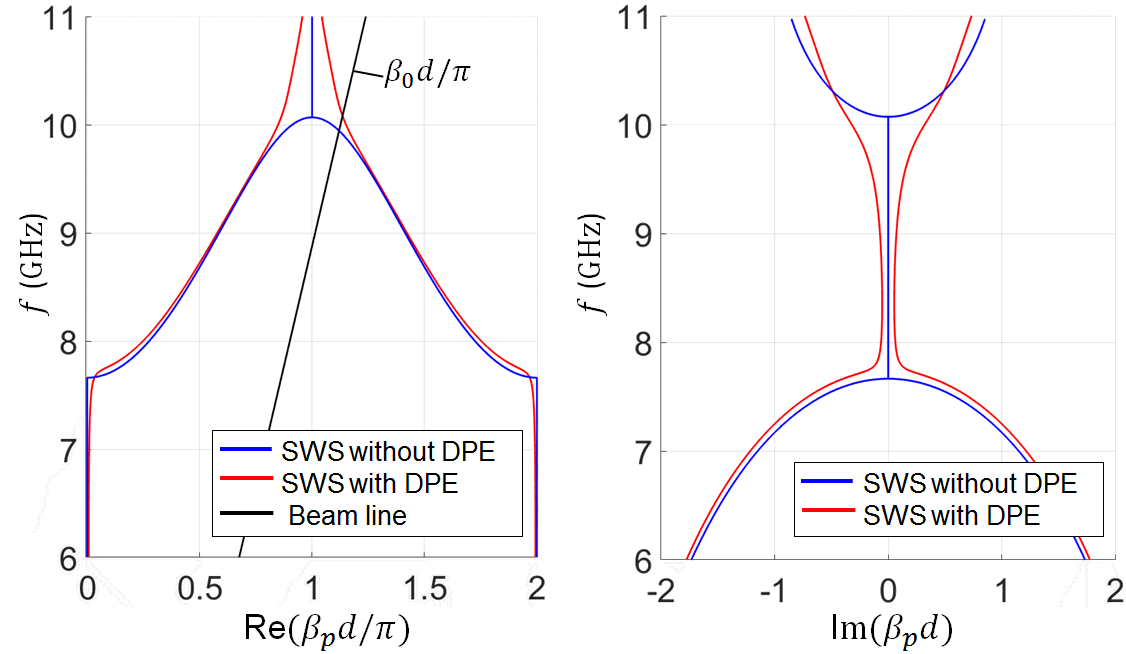}}
\par\end{centering}
\centering{}\caption{Details of the longitudinal cross-sections of a SWS without (a) and
with DPE (b). (c) Dispersion of EM guided modes in the ``cold''
SWSs in (a) and (b), without (blue curve) and with (red curve) distributed
power extraction (DPE), respectively. The dispersion shows the real
and imaginary parts of the complex wavenumber. The non-zero imaginary
part of wavenumber (red line) shows that the SWS in (b) exhibits distributed
power extraction. The black line is the ``beam line'' described
by $\beta_{0}=\omega/u_{0}$ , and the intersection point with the
curve of $\beta_{pr}=\mathrm{Re}(\beta_{pr})$ represents the approximative
syncronization point.}
\end{figure}

\begin{figure}
\begin{centering}
\centering \subfigure[]{\label{Fig:E_Field}\includegraphics[width=0.45\columnwidth]{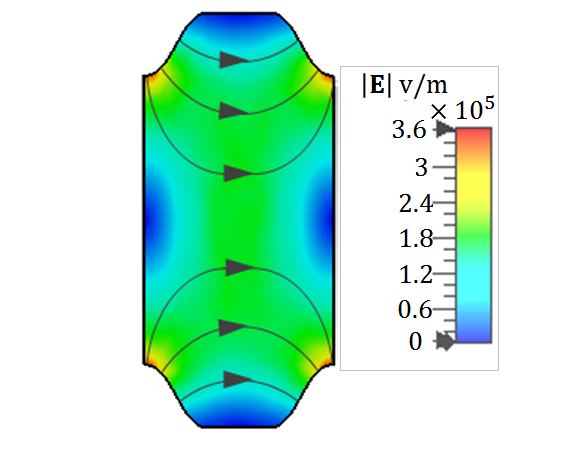}}\subfigure[]{\label{Fig:H_Field}\includegraphics[width=0.45\columnwidth]{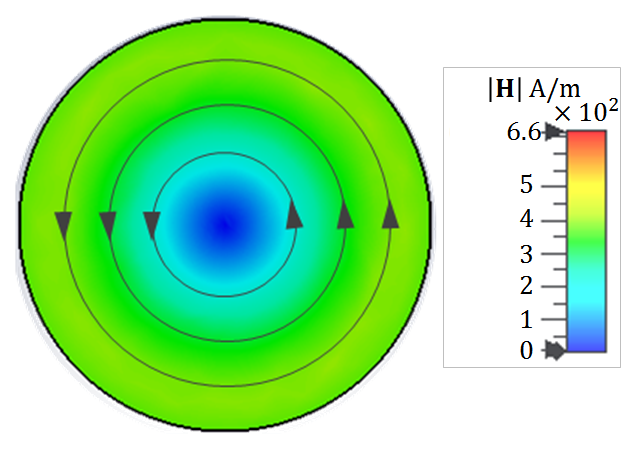}}
\par\end{centering}
\centering{}\caption{Field distribution for the TM-like mode supported by the SWS in Fig.
\ref{Fig:SWS_Conv-1}: (a) electric field on the longitudinal cross-section,
and (b) magnetic field on the transverse cross-section. Fields are
found with the mode solver of CST Studio Suite.}
\label{Fig:Fields}
\end{figure}

The DPE is introduced by adding two wire loops in each unit cell,
above and below as shown in Fig. \ref{Fig:SWS_EPD_BWO-1}, that couple
to the azimuthal magnetic field (shown in Fig. \ref{Fig:H_Field}),
and by Farady's Law an electromotive force is generated that excites
each coaxial waveguide, similarly to the way power is extracted from
magnetrons (Ch. 10 in Ref. \cite{gilmour1994principles}). The coaxial
cables have outer and inner radii equal to $2.57$ mm and $0.5$ mm,
respectively, leading to a 98 ohm characteristic impedance. Figure
\ref{Fig:Cold_Disp} shows a comparison between the dispersion relation
of the EMmodes in the two ``cold'' SWSs: one used in the conventional
BWO in Fig. \ref{Fig:SWS_Conv-1}, and the other one used in the BWO
with DPE in Fig. \ref{Fig:SWS_EPD_BWO-1}. The dispersion curves show
only the EM mode that is TM-like, i.e., the one with an axial (longitudinal)
electric field component, with electric and magnetic field distributions
shown in \ref{Fig:Fields}. The dispersion curves in Fig \ref{Fig:Cold_Disp}
show that the EM mode in the cold SWS with DPE is a backward wave
that has a propagation constant with non-zero imaginary part $\beta_{pi}$
at the frequency where the interaction with the e-beam occurs, i.e.,
at the point where the EM wave phase velocity $\omega/\beta_{pr}$
is synchronized to the velocity of electrons $u_{0}=0.88c$, where
$c$ is the speed of light in vacuum. This means that the cold SWS
in Fig. \ref{Fig:SWS_EPD_BWO-1} is suitable for our design of a BWO
with an EPD \cite{mealy2019backward,mealy2019exceptional}. An example
of the dispersion of the complex-wavenumber modes in the interactive
(``hot'') EM e-beam system with DPE has been shown in \cite{mealy2019backward,mealy2019exceptional}
using the Pierce-based model revealing the occurrence of an EPD. The
complex wavenumber dispersion relation in presence of DPE, shown in
Fig \ref{Fig:Cold_Disp}, is obtained by using two multi-mode ports
at the begin and end of a SWS unit-cell where each port has 30 modes
(almost all evanescent) that sufficiently represent the first TM-like
Floquet mode in the periodic SWS, while all the coaxial waveguides
are matched to their characteristic impedance to absorb all the outgoing
power. This is done using the Finite Element Frequency Domain solver
implemented in CST Studio Suite by DS SIMULIA that calculates the
scattering parameters of the unit cell, that have then been converted
to a transfer matrix to get the SWS complex Floquet-Bloch modes following
the same method in \cite{othman2016theory}.

\subsection{Hot simulations - oscillation frequency and fields}

\begin{figure}
\begin{centering}
\centering \subfigure[]{\label{Fig:Output_Conv-1}\includegraphics[width=0.9\columnwidth]{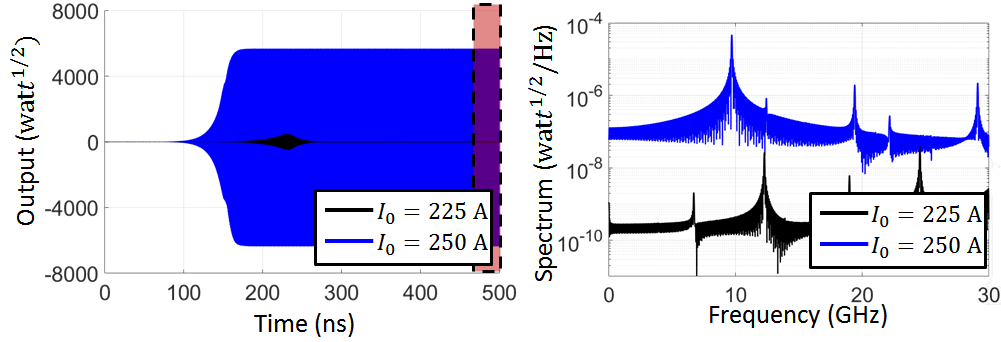}}
\par\end{centering}
\begin{centering}
\centering \subfigure[]{\label{Fig:Output_EPD-1}\includegraphics[width=0.9\columnwidth]{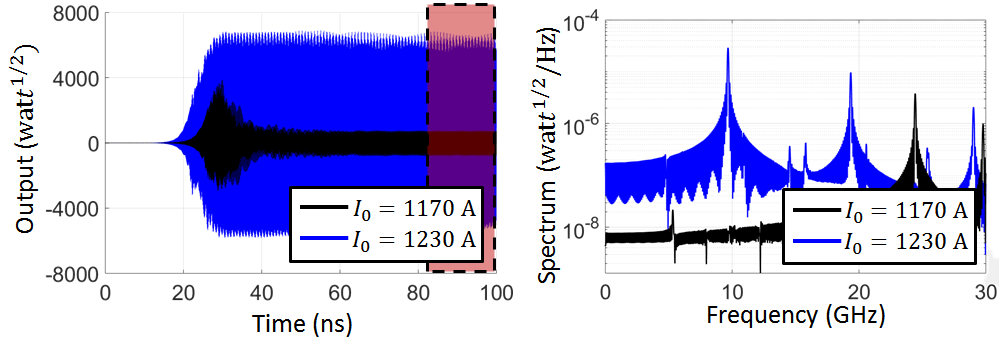}}
\par\end{centering}
\centering{}\caption{Output signal at the right-end waveguide port and its corresponding
spectrum when the SWS has 11 unit-cells, at (blue) and below (black)
the e-beam starting current for: (a) Conventional BWO and (b) EPD-BWO.
The frequency spectrum shows that there is not self-standing oscillation
at 9.7 GHz when the e-beam dc current is below the oscillation threshold,
i.e., when the current is below $250$A for theconventional BWO, and
below $1230$A for the EPD-BWO, but there is at these two e-beam current
values, hence they are the starting currents for the two types of
BWOs. It is important to stress that the figure shows only the ouput
power at the right-end port of the EPD-BWO, and indeed the output
value of the EPD-BWO is comparable to the one coming out of the conventional
BWO. The \textit{total} amount of power coming out of the EPD-BWO
is much higher than the one of the conventional BWO when we consider
all the other distributed ports, as shown in Fig. 5 of the main body
of the paper. }
\label{Fig:Output_Vs_Time-1}
\end{figure}

Simulations based on the particle-in-cell (PIC) solver, implemented
in CST Studio Suite, based on a relativistic annular e-beam with dc
voltage of $V_{0}=600$ kV, inner and outer radii of $R_{ib}=9$ mm
and $R_{ob}=10.3$ mm, respectively, and with dc axial magnetic field
of 2.6 T to confine the electron beam. The cathode is modeled using
the dc emission model with 528 uniform emission points. The full-wave
simulation uses around 1.3M Hexahedral mesh cells to model the SWS.

We study the starting e-beam current for oscillation in both types
of BWO (the conventional one, and the EPD-BWO in Fig. \ref{Fig:BWOs}
by sweeping the e-beam current $I_{0}$ and monitoring the RF power
and its spectrum of the waveguide output signal at the right end of
the cylindrical waveguide. Using a SWS with 11 unit-cells we show
in Fig. \ref{Fig:Output_Vs_Time-1} the output power at the main port
at the right end of the SWS when the e-beam current is just below
and just above the threshold current. A self-standing oscillation
frequency of 9.7 GHz is observed when the e-beam dc current $I_{0}$
for the conventional BWO is at or larger than than $250$A, while
for the EPD-BWO, self-standing oscillations is observed for an e-beam
current $I_{0}$ equal or greater than $1230$A. Such oscillations
are not observed for smaller e-beam current, as for example $225$A
for the conventional BWO and $1170$A for the EPD-BWO. Therefore we
conclude that the the starting current of oscillation is approximately
$250$A for the conventional BWO, and $1230$A for the EPD-BWO, when
the SWS length is 11 periods.

\begin{figure}
\begin{centering}
\centering \subfigure[]{\label{Fig:E_Field-1}\includegraphics[width=0.9\columnwidth]{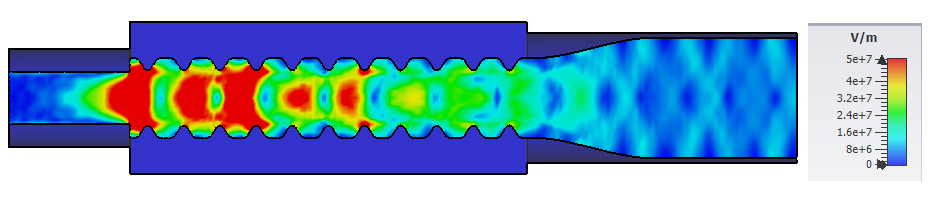}}
\par\end{centering}
\begin{centering}
\centering\subfigure[]{\label{Fig:E_Field_DPE}\includegraphics[width=0.9\columnwidth]{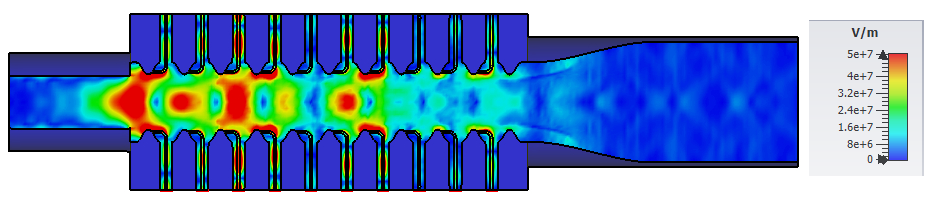}}
\par\end{centering}
\centering{}\caption{Electric field distribution in the SWS for: (a) conventional BWO and
(b) EPD-BWO. The figure in (b) shows power extraction in distributed
fashion from the coaxial waveguides.}
\label{Fig:Fields_BWOs}
\end{figure}

Figure \ref{Fig:Fields_BWOs} shows the electric field distribution
for the conventional BWO and the EPD-BWO when the e-beam dc current
$I_{0}$ is $1750$ A, in both cases, for a SWS of 11 unit cells.
The figure shows that ffor the conventional BWO the power is extracted
only from the main port at the right end, whereas for EPD-BWO most
of the power is extracted in a distributed fashion from the top and
bottom coaxial waveguides, resulting in much high power and high efficiency
as demonstrated in the main body of this paper.

\bibliographystyle{ieeetr}
\bibliography{myref}

\end{document}